\documentclass[12pt,epsf,epsfig,psfig]{article}
\usepackage{graphicx}
\usepackage{epsfig}
\usepackage{epstopdf}
\def\J{$J/\psi$}
\def\P{$\psi'$}
\def\X{$\chi_c$}
\def\j{J/\psi}
\def\X{$\chi_c$}

\def\P{$\psi'$}
\def\p{\psi'}
\def\U{$\Upsilon$}

\def\C{c{\bar c}}

\def\Q{Q{\bar Q}}
\def\e{\epsilon}
\def\t{\tau}

\def\NP{{ Nucl.\ Phys.\ }}
\def\PL{{ Phys.\ Lett.\ }}
\def\PR{{ Phys.\ Rev.\ }}

\def\PRL{{ Phys.\ Rev.\ Lett.\ }}

\def\ZP{{ Z.\ Phys.\ }}
\def\EPJ{{Eur.\ Phys.\ J.\ }}

\def\be{\begin{equation}}
\def\ee{\end{equation}}
\def\lsim{\raise0.3ex\hbox{$<$\kern-0.75em\raise-1.1ex\hbox{$\sim$}}}
\def\gsim{\raise0.3ex\hbox{$>$\kern-0.75em\raise-1.1ex\hbox{$\sim$}}}

\begin{document}
\begin{flushright}
BNL-NT-05/50\\
February 27, 2005
\end{flushright}

\vskip 1cm

\centerline{\Large \bf Sequential Charmonium Dissociation}

\vskip1cm

\centerline{\large  F.\ Karsch$^{a,b}$, 
D.\ Kharzeev$^{a}$ and H.\ Satz$^{b,c}$ } 

\vskip1cm

\begin{center}

a) Physics Department, \\ 
Brookhaven National Laboratory, \\ Upton, New York 11973-5000, USA\\
 
\vskip0.3cm
                    
b) Fakult{\"a}t f{\"u}r Physik, Universit{\"a}t Bielefeld, \\
Postfach 100 131, D-33501 Bielefeld, Germany\\
    
\vskip0.3cm
                    
c) Centro de F\'{\i}sica Te\'{o}rica de Part\'{\i}culas  (CFTP)\\
Instituto Superior T{\'e}cnico, Av. Rovisco Pais, \\ P-1049-001 Lisboa, 
Portugal
                    
\end{center}

\vskip 0.7cm

\centerline{\bf Abstract:}

\vskip0.5cm

Finite temperature lattice QCD indicates that the charmonium ground state
\J~can survive in a quark-gluon plasma up to 1.5 $T_c$ or more, while
the excited states \X~and \P~are dissociated just above $T_c$. We assume
that the \X~suffers the same form of suppression as that observed for
the \P~in SPS experiments, and that the directly produced \J~is unaffected 
at presently available energy densities. This provides a parameter-free 
description of \J~and \P~suppression which agrees quite well with that 
observed in SPS and RHIC data.

\vskip1cm

Recent studies of the behavior of charmonium states in a deconfined medium
show that the ground state \J(1S) survives up to considerably higher
temperatures than initially expected. In quenched QCD
\cite{Asakawa}-\cite{Iida}, charmonium correlators show no signs of 
medium-induced suppression at least up to 1.5 $T_c$, while above 
2 - 2.5 $T_c$, 
the signal is strongly modified or disappears. First work in QCD with 
two quark flavors supports these results \cite{Morrin}. In contrast, 
the higher excited states seem to disappear very near $T_c$; in quenched 
calculations, no signal for the \X~is seen at $T=1.1~T_c$ \cite{Datta}. 

\medskip

The results of direct spectral function studies are further supported
by potential model analyses \cite{Wong1}-\cite{Digal}, using the 
color-singlet free energy obtained in (quenched as well as unquenched)
lattice QCD to determine the heavy quark potential. These also lead to a 
\J~dissociation temperature of 2 $T_c$ or higher, while \X~and \P~disappear 
in the vicinity of 1.1 $T_c$.
In contrast, earlier potential model work \cite{K-M-S}-\cite{D-P-S1}, 
based on a heavy quark interaction which underestimated the actual $\Q$ 
potential, had predicted a considerably lower \J~dissociation temperature. 

\medskip

Since \J~suppression was proposed as a signature for quark-gluon plasma
formation in nuclear collisions \cite{M-S}, this modification of our
understanding of the in-medium behavior of charmonia can be quite
important for the interpretation of relativistic heavy ion data. 
Lattice calculations show that a temperature of 1.5 $T_c$ corresponds
to an energy density around 10 GeV/fm$^3$, and 2 $T_c$ to around
30 GeV/fm$^3$, which could move the suppression of direct \J~production
out of the range of RHIC.

\medskip

In hadron-hadron collisions \cite{Antoniazzi} it is found that about
60\% of the observed \J's are directly produced as (1S) states, with the
remainder coming to about 30\% from \X~and 10\% from \P~decay. The
hierarchy of suppression temperatures thus leads to a sequential
suppression pattern \cite{K-S,Gupta}, with an early suppression of 
the \P~and \X~decay products and a much later one for the direct 
\J~production.

\medskip

In this note, we want to consider the experimental results available now 
from the SPS and from RHIC, and show that the new theoretical understanding 
can be used to formulate a rather natural parameter-free description of 
the essential features of the data. 

\medskip

Our considerations are based on the following scenario. The \J~survival
probability $S_{\j}$ in $A-A$ collisions is defined as the ratio of 
the measured rate to that expected if the only modifications are due to 
the presence of normal nuclear matter. We assume that $S_{\j}$ consists
of one term $S_{\psi}$ corresponding to the survival of directly produced 
\J's and a second term $S_x$ for those coming from the decay of the higher 
excited states \X~and \P,
\be
S_{\j} = 0.6~S_{\psi} + 0.4~S_x.
\label{model}
\ee
The relative contributions here are those observed in hadron-hadron
collisions \cite{Antoniazzi}.
From the mentioned QCD studies we expect $S_{\psi}\simeq 1$
for energy densities up to 10 GeV/fm$^3$ or more, while $S_x$ is
expected to show suppression effects around the deconfinement point, i.e.,
for $\e \simeq 0.5 - 1.5$ GeV/fm$^3$. In principle, $S_x$ could
consist of two distinct terms, with different dissociation onsets for
\X~and \P. At present, however, neither calculational nor experimental
accuracy seems to permit such fine structure studies, and we shall 
therefore combine the decay of the two states into one term.

\medskip

We first turn to the onset pattern of suppression and consider the SPS 
data for \J~production from $Pb\!-\!Pb$ \cite{NA50EPJ} and $In\!-\!In$ 
interactions \cite{Roberta}, together with \P~data from $Pb\!-\!Pb$ 
collisions \cite{Sitta}; the analysis of \P~production in the $In\!-\!In$ 
data is not yet completed\footnote{We 
restrict ourselves here to symmetric ($A\!-\!A$) data and comment on the 
$S-U$ results later on.}. In addition, there are reference data from 
$p-\!A$ collisions with several nuclear targets \cite{Bordalo}, which 
define the necessary 
baseline for modifications of the production due to normal nuclear matter. 
The combined effect of all possible modifications was here parametrized 
in the form of nuclear absorption, leading to the absorption cross sections  
\be
\sigma_{\j}=4.3 \pm 0.3 ~{\rm mb}
\ee
for the \J~and 
\be
\sigma_{\p}=7.1 \pm 1.6~{\rm mb},
\ee 
for the \P, respectively \cite{Bordalo}. Using these in a Glauber 
analysis of $A-A$ data provides the production rates $(d\sigma_i/dy)_G$, with 
$i=\j,~\p$, as they would be if there were no effects beyond those caused 
by the presence of normal nuclear matter \cite{K-L-N-S}. The survival 
probability is then defined as
\be
S_i = {(d\sigma_i/dy) \over (d\sigma_i/dy)_G},
\label{surv}
\ee
describing whatever anomalous effects arise.  The centrality dependence of 
the $A\!-\!A$ data is determined through the number $N_{part}$ of 
participants, 
which is measured directly through a zero degree calorimeter. A Glauber
analysis then provides the density $n_{part}$ of participants in the 
transverse
overlap region $A$ of the collision \cite{K-L-N-S}, and the corresponding 
energy density is given by the Bjorken estimate 
\be
\e = {w_h \over A \tau_0} \left({dN_h \over dy}\right)_{AA}
= {\nu_h w_h \over \tau_0} n_{part}; 
\label{bj}
\ee
here $(dN_h/dy)_{AA}$ denotes the hadron multiplicity at the given 
centrality, $w_h$ the average hadron energy, and $\nu_h$ the average 
number of hadrons emitted per participant nucleon (the values of $\nu_h$ 
and $w_h$ can depend on centrality). For the equilibration time of the 
medium, we take $\tau_0=1$ fm, so that corrections for other possible 
values can easily be carried out. In our context, however, the 
formation time of the charmonium states in question should be less than
the formation time of the medium, which is the case if $\tau_0=1$ fm. 
The actual values of $\e$ we will cite
here were obtained by an event generator determination of the NA60 
collaboration and is based on VENUS \cite{Roberta}. It should be noted, 
however, that with constant $\nu_h\simeq 2$ and $w_h\simeq 0.5$ GeV, we 
get very similar results, while an event generator determination based on RQMD 
as input (used by the NA50 collaboration \cite{NA50EPJ}) leads to values 
which are higher by about 10\%.

\medskip

\begin{figure}[htb]
\centerline{\epsfig{file=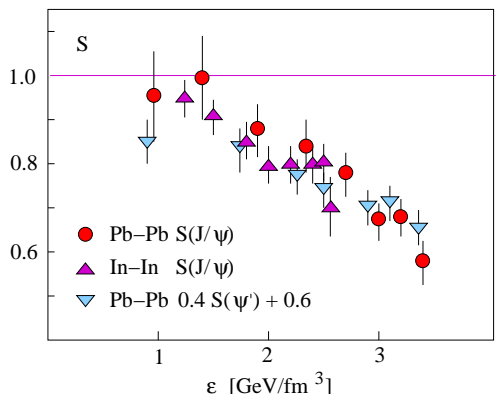,width=6.5cm}}
\caption{Universal \P~and \J~suppression at the SPS}
\label{scaling}
\end{figure}

We now return to our basic scenario, assuming that at present energy 
densities the directly produced \J~are unaffected, and the suppression 
patterns of the excited states \X~and \P~are about the same. This implies
that if we use the \P~data to form $0.4~S_{\p} + 0.6$, then as function
of the energy density this should coincide with the measured \J~results. 
In Fig.\ \ref{scaling}, we see that this is indeed quite well fulfilled,
for the overlap of \J~and \P~data as well as for the convergence to the
\J~``saturation'' value of about 60\%. 

\medskip

\begin{figure}[htb]
\centerline{\epsfig{file=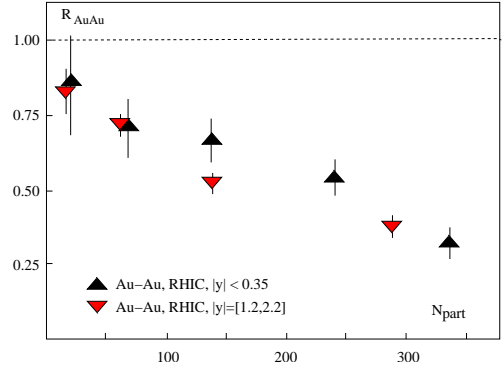,width=6.5cm}}
\caption{\J~production rates for $Au-\!Au$ collisions at 
$\sqrt s=200$ GeV \cite{hugo}}
\label{npart}
\end{figure}

Next we want to check if this pattern continues for higher energy densities
and therefore turn to the recently presented preliminary RHIC data; its
higher collision energy can provide correspondingly higher energy densities.
The \J~production rate $R_{Au-Au}$ in $Au\!-\!Au$ interactions is given 
relative to the result from scaled $p\!-\!p$ collisions, as shown in Fig.\  
\ref{npart} as function of the number of participant nucleons \cite{hugo}. 

\medskip

\begin{figure}[htb]
\centerline{\epsfig{file=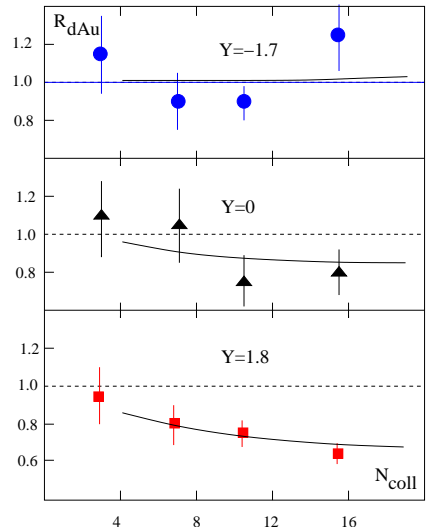,width=5.5cm}}
\caption{\J~production in $d-Au$ collisions at $\sqrt s=200$ GeV}
\label{dAu}
\end{figure}

In order to convert the rates $R_{Au-Au}$ into survival probabilities, we 
have to know what would be expected if only normal nuclear matter were 
present. At RHIC, this information is provided through $d\!-\!Au$ studies 
\cite{dAu}; the resulting nuclear modification factor, specifying the 
production rate relative to scaled $p\!-\!p$ collisions, is shown in 
Fig.\ \ref{dAu}.   

\medskip

To quantify these RHIC results, with their presently rather limited
statistics, we adopt a description similar to that used for SPS results 
and apply the well-known simplified absorption form
\be
S \simeq \exp\{-n_0 \sigma_{\rm diss} L\},
\label{path}
\ee
where $L$ denotes the path of the $\C$ in the nuclear medium and 
$n_0=0.17$ fm$^{-3}$ denotes normal
nuclear density. A Glauber analysis \cite{K-L-N} provides the relation 
between impact parameter $b$ and the number of collisions $N_{coll}$, and 
simple geometry gives $L =[R_A^2 - b^2]^{1/2}$ in terms of $b$ and the
nuclear radius $R_A$. A fit of Eq.\ (\ref{path}) to the data of 
Fig.\ \ref{dAu} gives\footnote{In the fit, we neglect the most 
peripheral point at $N_{coll}$, which corresponds to $b > R_{Au}$
and is thus due to nuclear surface rather than medium effects.}  
$$
\sigma_{\rm diss}(y=1.8) = 3.1 \pm 0.2 ~{\rm mb}
$$
\vskip-0.6cm
$$
\sigma_{\rm diss}(y=0) = 1.2 \pm 0.4 ~{\rm mb}
$$

\vskip-0.4cm
\be
\sigma_{\rm diss}(y=-1.7) = -0.1 \pm 0.2 ~{\rm mb}
\label{cross}
\ee

\medskip

for the corresponding \J~dissociation cross sections; for $y=-1.7$,
there are thus essentially no nuclear modifications. We note that
here, as for the SPS case, these cross sections are just a global
way to account for whatever nuclear effects can arise. A more detailed
analysis based on shadowing and absorption is given in \cite{ramona-shadow}; 
an analysis based on the Color Glass Condensate approach has recently been 
performed in \cite{Kharzeev:2005zr}. In the latter approach, 
the factorization of the shadowing and absorption corrections does not occur; 
nevertheless, here we use the equation (\ref{path}) just as a way to 
parameterize the data.

\medskip

\begin{figure}[htb]
\centerline{\epsfig{file=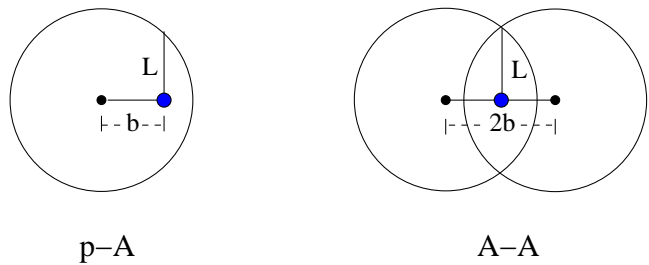,width=8.5cm}}
\caption{Impact parameter relation between $p-\!A$ and $A-\!A$ collisions}
\label{impact}
\end{figure}

For $A\!-\!A$ collisions at RHIC energy, we make use of the same simplified 
form (\ref{path}). The geometry connecting the impact parameter $b$ and path 
length $L$ in $p-Au$ and $Au-Au$ collisions is illustrated in Fig.\ 
\ref{impact}; the relation between $b$ and $N_{part}$ is again given by 
a Glauber analysis \cite{K-N}. We thus here obtain for the survival 
probability 
\be
S^{AA}_i(y,N_{part})= {R_{AA}(y,N_{part}) \over 
\exp\{-n_0 [\sigma_{\rm diss}(y)
+ \sigma_{\rm diss}(-y)] L\}},
\label{S-AA}
\ee
corresponding to the fact that for $y \not= 0$ the charmonium state
passes one nucleus at rapidity $y$, the other at rapidity $-y$.

\medskip

\begin{figure}[htb]
\vspace*{0.2cm}
\centerline{\epsfig{file=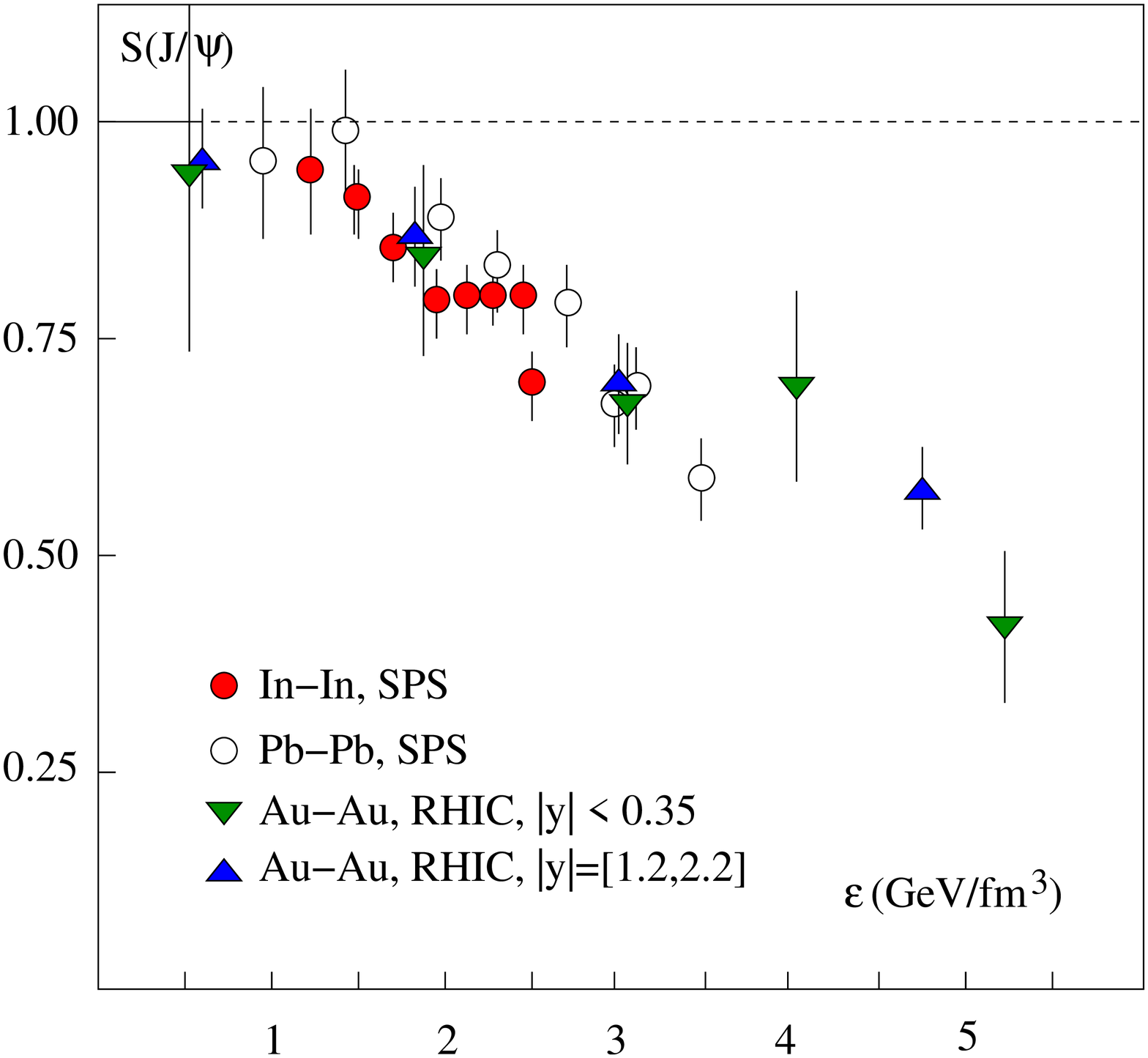,width=7cm,height=5cm}}
\caption{\J~suppression as function of energy density}
\label{in-rhic}
\end{figure}

Applying eq.\ \ref{S-AA} to the rates shown in Fig.\ \ref{npart} together
with the nuclear modification cross sections (\ref{cross}) provides the
survival probability as function of $N_{part}$. The corresponding energy
densities have been calculated in a Glauber analysis based directly on
the PHENIX $E_T$ data \cite{phenix-bj}, and in Fig.\ \ref{in-rhic} we
compare the RHIC results to those from the SPS.

\medskip

It is seen that the two data sets are quite compatible, both in the onset 
and in the flattening at about 50 - 60\%. Concerning the RHIC data, it 
should be emphasized that the choice of $\t_0=1$ fm is certainly debatable; 
a smaller value would move the RHIC points to correspondingly larger $\e$ 
values.
%\footnote{It should be noted that we are interested in the energy 
%density at the time when the quarkonium state is formed, and this formation 
%time should not depend on the collision energy.}.

\medskip

So far we have considered only symmetric ($A\!-\!A$) collisions. We find,
however, that the \P~production measured in $S\!-\!U$ interactions at the SPS
\cite{SU} also agrees quite well with the pattern shown in Fig.\ 
\ref{scaling}. In contrast, the reported $S\!-\!U$ \J~rates \cite{SU}
do not show an onset of suppression 
at the centrality at which it sets in for $In\!-\!In$ collisions. The reason 
for this is not clear, although two special features have been pointed out. 
The centrality dependence of the $S\!-\!U$ data is determined by transverse
energy ($E_T$) measurements, not by the more reliable method based on
the zero degree calorimeter specifying directly the number of spectator
nucleons. For $Pb-Pb$ collisions, it is observed that the centrality
dependences obtained from $E_T$ and $E_{ZDC}$ measurements can in fact
show differences. Moreover, it has been noted that a 10\% shift in the 
normalization of the $S\!-\!U$ data would lead to full agreement between 
all SPS data sets.

\medskip

 \J~production at RHIC has also been adressed in terms of anomalous 
suppression followed by regeneration at hadronization \cite{recom}. Such 
a scenario assumes first a strong anomalous suppression of the overall 
\J~production, including that of the $1S$ state, and subsequently a 
renewed \J~formation at the hadronization stage, due to a pairing 
of $c$ and$\bar c$ quarks from different nucleon-nucleon collisions. 
The latter mechanism becomes possible at RHIC energies because of abundant 
$\C$ production. It leads to rates increasing with centrality, which 
are taken to just compensate the dropping direct production. In such 
an approach, the agreement between central SPS data (with no regeneration) 
and RHIC rates (with considerable regeneration) is coincidental. 
We also note that the anomalous suppression assumed for direct \J~production 
in the regeneration approach is not in accord with what we know today about 
\J~survival in a quark-gluon plasma, as found in statistical QCD.

\medskip

Finally we turn to a further check of these considerations. It was
pointed out some time ago that the effect of \J~suppression could  
also manifest itself in the transverse momentum behaviour 
\cite{early,K-N-S}, and in fact the pattern resulting from
sequential decay differs strongly from that due to 
regeneration \cite{T-M}.

\medskip

The basic effect of a nuclear medium on the transverse momentum
behaviour of hard processes is a collision broadening of the incident 
parton momentum; this in turn leads to a broadening of the transverse
momentum distribution of the charmonia formed by hard parton interactions,
(dominantly gluon fusion). 
It was shown that a random walk approach leads to an average squared 
transverse \J~momentum  
\be
\langle p_T^2 \rangle_{pA} = \langle p_T^2 \rangle_{pp} + N_c^A \delta_0
\label{pTpA}
\ee
for $p\!-\!A$ and to
\be
\langle p_T^2 \rangle_{AA} = \langle p_T^2 \rangle_{pp} + N_c^{AA} \delta_0
\label{pTAA}
\ee
for $A\!-\!A$ collisions. Here $N_c^A$ denotes the average number of 
pre-fusion collisions of the projeticle parton in the target nucleus $A$, 
and $N_c^{AA}$ the sum of the average number of collisions of a projectile 
parton in the target and vice versa, at the given centrality. 
The parameter $\delta_0$ specifies the average ``kick'' which 
the incident parton receives in each subsequent collision. The basic 
parameters determining the $p_T$-broadening in nuclear matter 
are thus the elementary $\langle p_T^2 \rangle_{pp}$ from 
$p\!-\!p$ interactions and the value of $\delta_0$, determined by 
corresponding $p\!-\!A$ data; both depend on the collision energy. 
The $A$-dependence of $N_c^A$ as well as the behaviour of $N_c^{AA}$ 
as function of centrality can be obtained through a Glauber analysis; 
the latter defines the ``normal'' centrality dependence of 
$\langle p_T^2 \rangle_{AA}$. Such an analysis also has 
to include the normal absorption of the produced charmonia in nuclear 
matter; this effectively shifts the fusion point for the observed charmonia 
further ``down-stream'' \cite{K-N-S}.
 
\medskip

\begin{figure}[h]
\begin{minipage}[t]{7cm}
\epsfig{file=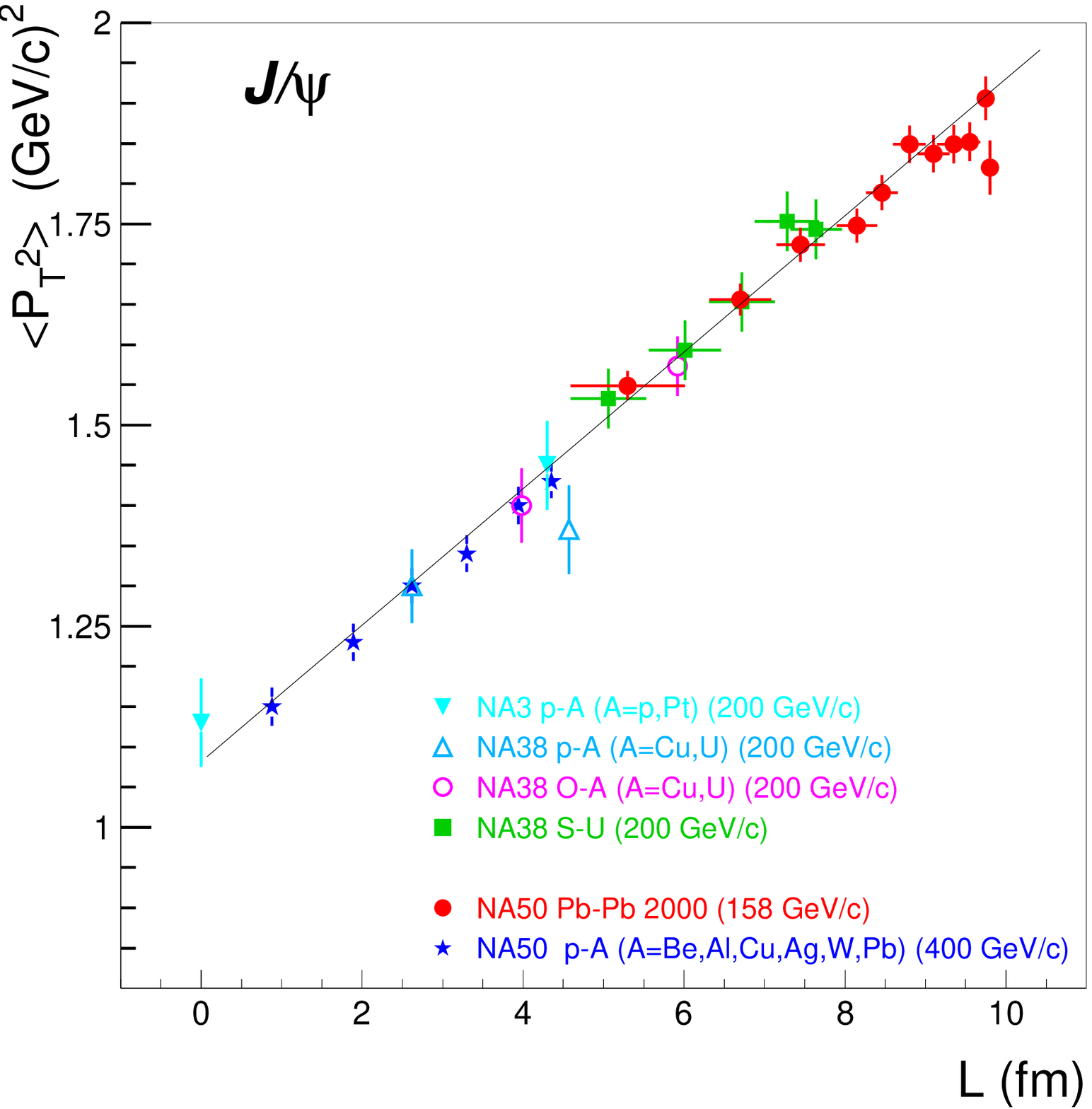,width=7.5cm,height=5.5cm}
\vspace*{-0.5cm}
\caption{\J~transverse momentum behaviour at the SPS\cite{transv}}
\label{PT}
\end{minipage}
\hspace{1.6cm}
\begin{minipage}[t]{7cm}
\vspace*{-4.9cm}
%\hskip-0.3cm
\epsfig{file=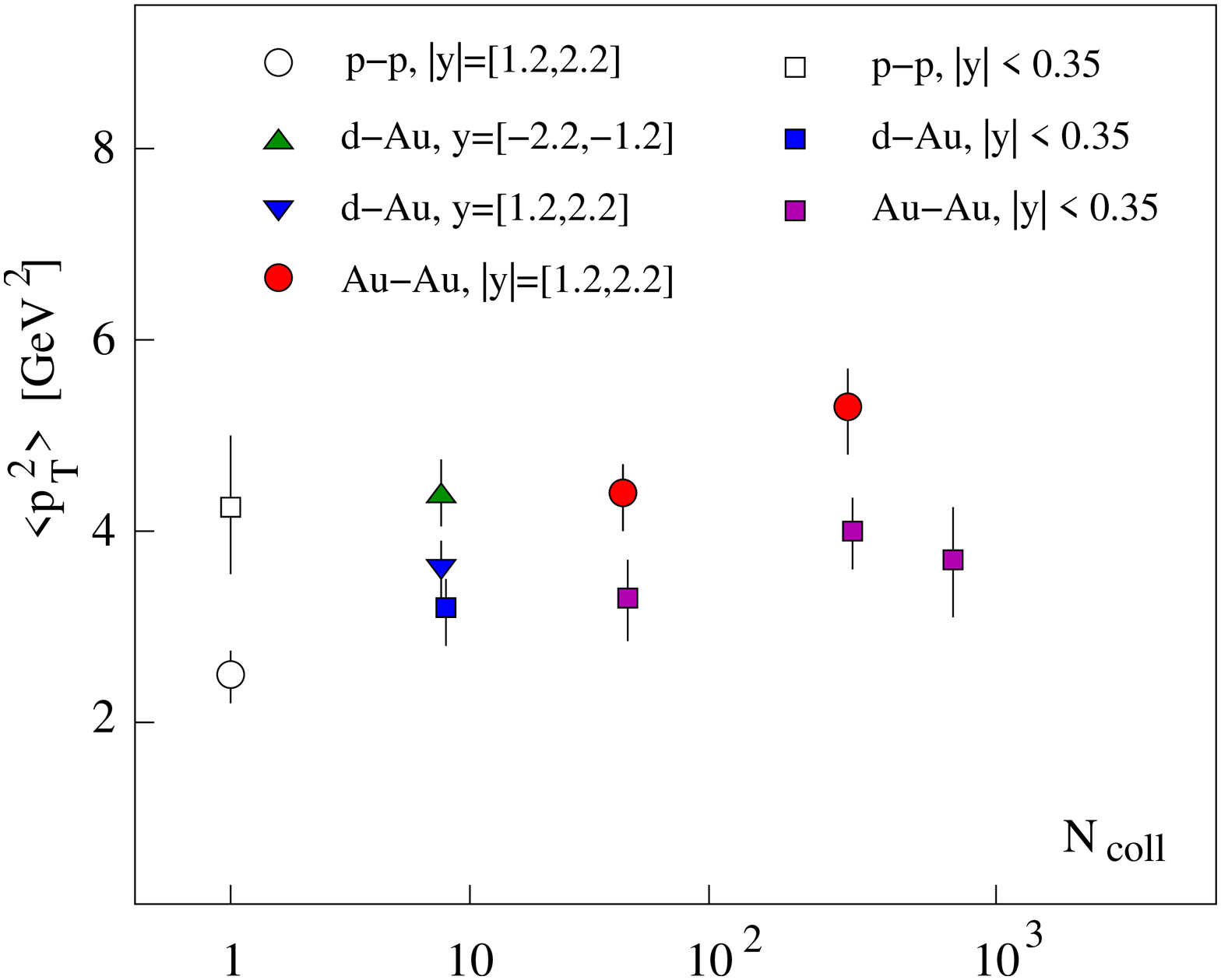,width=6.4cm,height=4.7cm}
\vspace*{0.2cm}
\caption{\J~transverse momentum behaviour at RHIC \cite{hugo,Gunji}}
\label{PT-rhic}
\end{minipage}
\end{figure}

A compilation of \J~transverse momentum data from the SPS \cite{transv}
is shown in Fig.\ \ref{PT}; it clearly indicates first the increase of the 
average transverse momentum from $p\!-\!p$ to $p\!-\!A$ (for $A=Pb$), and 
then a further increase with centrality for nucleus-nucleus collisions.
The preliminary data for the average $p_T^2$ observed in \J~production 
at RHIC is shown in Fig.\ \ref{PT-rhic} \cite{hugo,Gunji}. Here we note
that while the muon arm data ($ |y| \in [1.2,2.2]$) shows the expected 
broadening when going from $p\!-\!p$ to $d\!-\!Au$, the central electron
data ($|y|\leq 0.35$) does not follow this pattern. Since our analysis is
based on such a broadening, we concentrate here on muon data. More statistics
at central rapidity should clarify this problem. We note that since 
both $\langle p_T^2 \rangle_{pp}$ and $\langle p_T^2 \rangle_{dA}$
can in general depend on rapidity as well as on collision energy, each
data set requires a separate analyis. 

\medskip

At SPS energy, one has $\langle p_T^2 \rangle_{pp} = 1.25 \pm 0.05$ 
(GeV/c)$^2$ and $\langle p_T^2 \rangle_{pU} = 1.49 \pm 0.05$ (GeV/c)$^2$ 
\cite{transv}. The average number of pre-fusion collisions is calculated
in a Glauber analysis \cite{K-N-S}, and with the normal nuclear absorption 
specified by the average of eqns.\ (2/3) it is found to be about 3. From 
eq.\ (\ref{pTpA}) we then obtain
\be
\delta_0^{SPS} = 0.083 \pm 0.023~{\rm GeV}^2
\label{delta-sps}
\ee 
for the average projectile parton broadening in the target nucleus.

\medskip

From the RHIC $\mu^+\mu^-$ data, we obtain $<p_T^2>_{pp} = 2.51 \pm 0.21$ 
(GeV/c)$^2$ and $<p_T^2>_{dAu} = 3.96 \pm 0.28$ (GeV/c)$^2$ \cite{hugo,Gunji}; 
for the latter value, we have taken the average of the positive and negative 
rapidity ranges, since this is also done for the corresponding $Au\!-\!Au$ 
data. A corresponding Glauber analysis, with normal nuclear absorption 
specified by eq.\ (\ref{cross}), gives nearly 3.5 pre-fusion parton collisions 
and leads to
\be
\delta_0^{RHIC} = 0.42 \pm 0.09~{\rm GeV}^2
\label{delta-rhic}
\ee 
for the corresponding parton broadening in the large rapidity region.

\medskip

\begin{figure}[htb]
\centerline{\epsfig{file=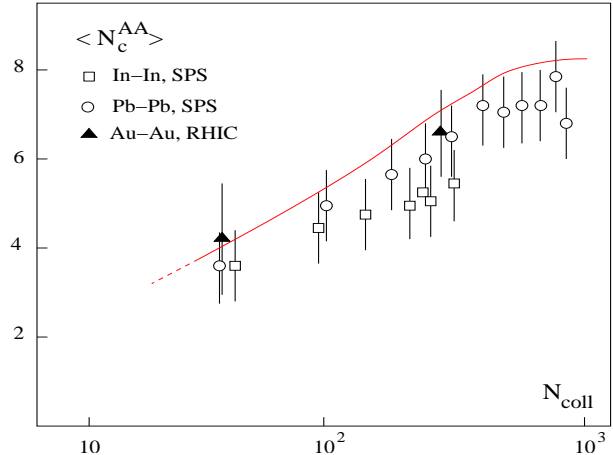,width=8cm,height=6cm}}
\caption{Transverse momentum broadening at SPS and RHIC}
\label{pTcurve}
\end{figure}

{In a sequential dissociation scenario, the transverse momentum behaviour
below the onset of exited state suppression is that of charmonia suffering 
only initial state broadening and normal nuclear absorption. Once the
higher states are suppressed, one has once again only direct \J's
experiencing initial state effects and normal absorption. Hence
apart from possible fluctuations in the suppression region, one should 
observe the $p_T$ behaviour as given by eqs.\ (\ref{pTpA}) and (\ref{pTAA}). 
In other words, the \J~transverse momentum should be determined only by 
the initial nuclear medium. This again predicts a common behaviour of 
measurements from SPS and RHIC. Given the values of $\delta_0$ as
determined above, data for  $\langle p_T^2 \rangle_{AA}$
and $\langle  p_T^2 \rangle_{pp}$ define  
\be
N_c^{AA}  = \{\langle p_T^2 \rangle_{AA} - 
\langle  p_T^2 \rangle_{pp}\} / \delta_0
\label{pTmeasure}
\ee
as a characteristic measure of transverse momentum behaviour.
In Fig.\ \ref{pTcurve} we show the SPS data from $Pb\!-\!Pb$ \cite{transv} 
and $In\!-\!In$ \cite{In-pT} collisions together with the RHIC muon data 
\cite{hugo,Gunji} and find that they indeed agree quite well. Once the
corresponding broadening pattern for the RHIC electron data is determined,
it should also follow this curve, even though the centrality dependent
values for $\langle p_T^2 \rangle_{AA}$ can be quite different.}

\medskip

As mentioned, the centrality dependence of $N_c^{AA}$ has also
been calculated directly in a Glauber analysis \cite{K-N-S}; the result 
is included in Fig.\ \ref{pTcurve}. It lies consistently somewhat higher 
than the results obtained from the data, which presumably comes from
using a larger normal suppression.

\medskip

In contrast to the increasing $p_T$-broadening determined 
by initial state parton scattering, \J~production through $\C$ pairing
at hadronization leads to a centrality-independent  
$\langle p_T^2 \rangle_{AA}$ \cite{T-M}. Remnant direct production will
of course modify this, but a strong regeneration component should in
any case considerably weaken the centrality dependence.

\medskip

The lack of the feed-down contributions to the observed \J~'s and 
the presence of the plasma may also affect \J~ polarization 
\cite{Ioffe:2003rd,Gupta:1998ut}, even though the theoretical 
description of quarkonium polarization has so far been notoriously 
difficult. Nevertheless, the predicted change of polarization may occur, 
and should be investigated experimentally. 

\medskip

We conclude that present \J~production data agree quite well with the
expectations based on quark-gluon plasma formation. The observed
onset of anomalous \J~suppression now coincides, within errors, with 
that found for \P~production, and the corresponding energy density
agrees with that expected from finite temperature QCD for the dissociation 
of higher excited charmonium states. The \J~production remaining beyond
this initial anomalous suppression, at about 60 \%, agrees with that 
predicted by a survival of directly produced $1S$ charmonium states and
thus is also in accord with present QCD calculations. Further checks can 
come from measurements of \P~production in $In\!-\!In$ collisions, 
from an eventual onset of direct \J~suppression at higher $\e$ (LHC),
and from similar results for \U~production in nuclear collisions.

\vskip1cm

\centerline{\bf Acknowledgements}

\vskip0.3cm

It is a pleasure to thank R.\ Arnaldi, P.\ Bordalo, R.\ Granier de Cassagnac, 
T.\ Gunji, C.\ Louren{\c c}o, M.\ Nardi and R.\ L.\ Thews for discussions and 
valuable help in obtaining and/or analyzing the data.  The work of D.K. and 
F.K. was supported by the U.S. Department of Energy under Grant No. 
DE-AC02-98CH10886.

\end{document}